\documentclass[nofootinbib,prd,showpacs]{revtex4-1}
\usepackage{amsmath}
\usepackage{amsfonts}
\usepackage{amssymb}
\usepackage{graphicx}
\usepackage{ulem}
\usepackage[usenames,dvipsnames]{xcolor}
\usepackage{hyperref}   
\hypersetup{colorlinks=true, 
citecolor=MidnightBlue, linkcolor=MidnightBlue, urlcolor=MidnightBlue}
\usepackage{tikz}
\definecolor{purple}{rgb}{1,0,1}
\definecolor{lime}{HTML}{A6CE39} % needs xcolor

\newcommand{\orcidicon}{%
	\begin{tikzpicture}
	\draw[lime, fill=lime] (0,0) 
		circle [radius=0.16] 
		node[white] {{\fontfamily{qag}\selectfont \tiny ID}};
	\draw[white, fill=white] (-0.0625,0.095) 
		circle [radius=0.007];
	\end{tikzpicture}	\hspace{-2mm}
}
\newcommand\orcidFrancisco{{\href{https://orcid.org/0000-0002-9388-8373}{\orcidicon}}}
\newcommand\orcidMatt{{\href{https://orcid.org/0000-0003-1088-6485}{\orcidicon}}}
\newcommand\orcidAlex{{\href{https://orcid.org/0000-0002-1763-3563}{\orcidicon}}}
\begin{document}
%=================================================================
\title{Dynamic thin-shell black-bounce traversable wormholes}
%=================================================================
\author{Francisco S. N. Lobo\orcidFrancisco\!\!}
\email{fslobo@fc.ul.pt}
\affiliation{Instituto de Astrofísica e Ci\^encias do Espa\c{c}o, Facultade de Ci\^encias da Universidade de Lisboa, Edif\'icio C8, Campo Grande, P-1749-016, Lisbon, Portugal}
%=================================================================
\author{Alex Simpson\orcidAlex\!\!}
\email{alex.simpson@sms.vuw.ac.nz}\affiliation{School of Mathematics and Statistics, Victoria University of Wellington, PO Box 600, Wellington 6140, New Zealand}
%=================================================================
\author{Matt Visser\orcidMatt\!\!}
\email{matt.visser@sms.vuw.ac.nz}\affiliation{School of Mathematics and Statistics, Victoria University of Wellington, PO Box 600, Wellington 6140, New Zealand}
%-----------------------------------------------------------------
%=================================================================
\date{23 March 2020; \LaTeX-ed \today}
%=================================================================
\begin{abstract}
%=================================================================

Based on the recently introduced black-bounce spacetimes, 
we shall consider the construction of the related spherically symmetric thin-shell traversable wormholes within the context of standard general relativity.
All of the really unusual physics is encoded in one simple parameter $a$ which characterizes the scale of the bounce. Keeping the discussion as close as possible to standard general relativity is the theorist's version of only adjusting one feature of the model at a time. 
We shall modify the standard thin-shell traversable wormhole construction, each bulk region now being a black-bounce spacetime, and with the physics of the thin shell being (as much as possible) derivable from the Einstein equations. Furthermore, we shall apply a dynamical analysis to the throat by considering linearized radial perturbations around static solutions, and demonstrate that the stability of the wormhole is equivalent to choosing suitable properties for the exotic material residing on the wormhole throat. The construction is sufficiently novel to be interesting, and sufficiently straightforward to be tractable. 


%=================================================================
\end{abstract}
%=================================================================
%=================================================================
\maketitle
%=================================================================
\def\d{{\mathrm{d}}}
%=================================================================
\tableofcontents
%=================================================================
\section{Introduction}
%=================================================================

One can tentatively trace back wormhole physics to Flamm's work in 1916 \cite{Flamm} and to the ``Einstein-Rosen bridge'' wormhole-type solutions considered by Einstein and Rosen (ER), in 1935 \cite{Einstein-Rosen}. However, the field lay dormant for approximately two decades until 1955, when John Wheeler became interested in topological issues in general relativity \cite{geons}. He considered multiply-connected spacetimes, where two widely separated regions were connected by a tunnel-like gravitational-electromagnetic entity, which he denoted as a ``geon''. These were hypothetical solutions to the coupled Einstein-Maxwell field equations.
Subsequently, isolated pieces of work do appear, such as the Homer Ellis' drainhole concept \cite{homerellis,homerellis2}, Bronnikov's tunnel-like solutions \cite{bronikovWH}, and Clement's 
five-dimensional axisymmetric regular multiwormhole solutions \cite{Clement:1983tu}, until the full-fledged renaissance of wormhole physics in 1988, through the seminal paper by Morris and Thorne \cite{Morris}.

In fact, the modern incarnation of Lorentzian wormholes (and specifically traversable wormholes) now has over 30 years of history. Early work dates from the late 1980s~\cite{Morris,Yurtsever, Visser:surgical,Visser:examples,Visser:1989a,Visser:1989b,Visser:1989c}. 
Lorentzian wormholes became considerably more mainstream in the 1990s~\cite{Visser:1990,Hochberg:1990,Frolov:1990,Visser:1990-PRD, Visser:1990-mpla, Visser:1992, Visser:1993, natural,Kar1,Kar2,Poisson,Visser:book,gr-qc/9704082,gr-qc/9710001,Visser:1996, Teo:1998,hochvisserPRL98,hochvisserPRD98,gr-qc/9901020, Hochberg:1998,gr-qc/9908029}, including work on energy condition violations~\cite{gvp1,gvp2,gvp3,gvp4,gvp:jerusalem,Visser:cosmo99,Martin-Moruno:2017}, with significant work continuing into the decades 2000-2009~\cite{Barcelo:2000, gr-qc/0003025, Dadhich:2001,gr-qc/0205066, VKD, gr-qc/0405103, gr-qc/0406083, Arellano:2006, Lobo:2007, Lobo-Nova,Boehmer:2007rm,Boehmer:2007md, modgravity1, modgravity2} and 2010-2019~\cite{modgravity3, modgravity4, Bohmer:2011, 
Montelongo-Garcia:2011, Lobo:2012, 
Harko:2013,Mehdizadeh:2015jra,Zangeneh:2015jda,Lobo:2015,
Boonserm:2018,Simpson:2018,Simpson:2019,Lobo:2017oab,Rosa:2018jwp}.
We shall particularly focus on the thin-shell formalism~\cite{Sen, Lanczos, Darmois, 
Synge, Lichnerowicz, Israel}, first applied to Lorentzian wormholes in~\cite{Visser:surgical, 
Visser:examples}, and subsequently further developed in that and other closely related settings by many other authors~\cite{Eiroa,Lobo-Crawford,FHK,BLP,Goncalves,LLQ,
Lobo-ec-dust, Lobo-CQG, LL-PRD, Sushkov, Lobo-phantom,Lobo-Crawford-2,
Lobo-phantom2, Eiroa:cylindrical, Eiroa:dilaton, Eiroa:gauss-bonnet, 
Rahaman1, Rahaman2, Eiroa:cosmic-strings, Eiroa:gas, Lemos:plane, Eiroa:brans, 
Eiroa:generalized-gas, Eiroa:stability, Eiroa:cylindrical2, Halisoy, Dias, Yue-Gao, 
Lobo:phantom-stars, Lake1, Lake2}.
Our notation will largely follow that of Hawking and Ellis~\cite{hawking-ellis}.
For the purposes of this article we will focus primarily on applying the technical machinery built up regarding spherically-symmetric thin-shell spacetimes in references~\cite{Visser:surgical,Visser:book,Montelongo-Garcia:2011,Lobo:2012,Lobo:2015}, and for the bulk spacetimes (away from the thin shell)
shall restrict attention to the recently developed black-bounce spacetimes of references~\cite{Simpson:2018,Simpson:2019} (for related work, we refer the reader to \cite{Bronnikov:2006fu,Bronnikov:2003gx}).

Consider the following candidate regular black hole (a black bounce) specified by the spacetime line element:
\begin{equation}
    ds^{2}=-\left(1-\frac{2m}{\sqrt{u^{2}+a^{2}}}\right)dt^{2}+
    \left(1-\frac{2m}{\sqrt{u^{2}+a^{2}}}\right)^{-1}\,du^2 +
    \left(u^{2}+a^{2}\right)\, d\Omega^{2} \,.
\end{equation}
Here the $u$ and $t$ coordinates have the domains $u\in(-\infty,+\infty)$ 
and $t\in(-\infty,+\infty)$.
In the original references~\cite{Simpson:2018,Simpson:2019} the $u$ coordinate was called $r$, however we now want to use the symbol $r$ for other purposes. 
In this work, we shall analyse thin-shell constructions based on this spacetime. By considering the coordinate transformation $r^2=u^2 + a^2$, we shall use the following completely equivalent line element:
\begin{equation}
    ds^{2}=-\left(1-\frac{2m}{r}\right)dt^{2}+
    \left(1-\frac{2m}{r}\right)^{-1}\left(1-\frac{a^2}{r^2}\right)^{-1}\,dr^2    
    +r^{2}\, d\Omega^{2}\,.
\end{equation}
Here the $r$ coordinate is now a double-cover of the $u$ coordinate. It has the domain $r\in(a,+\infty)$, and can be interpreted as the Schwarzschild area coordinate --- all the 2-spheres of constant $r$ have area $A(r)=4\pi r^2$.
This choice has the additional advantage of making it easy to directly compare the current analysis with most other work in the literature.
The  $t$ coordinate has the usual domain $t\in(-\infty,+\infty)$.
All of these black-bounce spacetimes are simple one-parameter modifications of Schwarzschild spacetime; for detailed analyses of the properties of these black-bounce spacetimes see references~\cite{Simpson:2018,Simpson:2019}.

While Lorentzian wormholes are in general very different from Mazur--Mottola gravastars~\cite{Mazur:2001,Mazur:2003, Mazur:2004a, Mazur:2004b,gravastar1,gravastar2,Lobo-gravastar1,Lobo-gravastar2}, it is worth pointing out that in the thin-shell approximation there are very many technical similarities --- quite often a thin-shell wormhole calculation can be modified to provide a thin-shell gravastar calculation at the cost of flipping a few strategic minus signs~\cite{Martin-Moruno:2011,Lobo:2015-novel-gravastar}.  

Structurally we organize the article as follows: We first introduce and briefly summarize  the appropriate variant of the thin-shell formalism in Section~\ref{S:formalism}. Section \ref{S:applications} discusses some specific examples and applications. Finally we conclude in Section~\ref{S:conclusion}.

%=================================================================
\section{Thin-shell formalism}\label{S:formalism}
%=================================================================
We shall first perform a general (and relatively straightforward) theoretical analysis,  somewhat along the lines laid out in  reference~\cite{Montelongo-Garcia:2011}, but with appropriate specializations, simplifications, and modifications. Subsequently we shall investigate a number of specific examples in the way of special cases and toy models.
We discuss the bulk spacetimes in subsection~\ref{SS:bulk}, the extrinsic curvature of the thin shells in subsection~\ref{SS:extrinsic}, before moving on to the Lanczos equations in subsection~\ref{SS:Lanczos}. We then discuss the Gauss and Codazzi equations in subsection~\ref{SS:Gauss}, before considering the equation of motion and its linearization in subsections~\ref{SS:eom} and~\ref{SS:linearized}. Finally we develop the master equation in subsection~\ref{SS:master}, before moving on to the next section~\ref{S:applications} where we shall discuss some specific examples and applications.

%###########################################################
\subsection{Bulk spacetimes}\label{SS:bulk}
%###########################################################

We initiate the discussion by considering two distinct ``bulk'' spacetime manifolds, ${\cal M_+}$ and ${\cal M_-}$, equipped 
with boundaries $\partial{\cal M_+}= \Sigma_+$  and $\partial{\cal M_-}= \Sigma_-$. As long as the boundaries are isometric, $ \Sigma_+\sim \Sigma_-$, then we can define a manifold ${\cal M} = {\cal M_+} \cup {\cal M_-}$, which is smooth except possibly for a thin-shell transition layer at  $ \Sigma_+\sim\Sigma_-$.
In particular, consider two  static spherically symmetric black-bounce spacetimes given on ${\cal M_\pm}$ by the following two-parameter $(m,a)$ Lorentzian-signature  line elements $g_{\mu \nu}^+(x^{\mu}_+)$ and $g_{\mu \nu}^-(x^{\mu}_-)$:
\begin{equation}
    ds^{2}=-\left(1-\frac{2m_{\pm}}{r_{\pm}}\right)dt_{\pm}^{2}+
    \left(1-\frac{2m_{\pm}}{r_{\pm}}\right)^{-1}\left(1-\frac{a_{\pm}^2}{r_{\pm}^2}\right)^{-1}\,dr_{\pm}^2    
    +r_{\pm}^{2}\, d\Omega_{\pm}^{2}\,.
    \label{generalmetric}
\end{equation}
The usual Einstein field equations, $G_{{\mu}{\nu}}=8\pi \,T_{{\mu}{\nu}}$ (with $c=G=1$), imply that the physically relevant 
orthonormal components of the stress-energy tensor are  (in the two bulk regions) specified by:
\begin{eqnarray}
\rho(r)&=&-\frac{1}{8\pi }\frac{a^2\left( r-4m \right)}{ r^5},\label{rho}\\
p_{r}(r)&=&-\frac{1}{8\pi }\frac{a^2}{ r^4},\label{pr}\\
p_{t}(r)&=&\frac{1}{8\pi }\frac{a^2\left( r-m \right)}{ r^5}\,.
\label{pt}
\end{eqnarray}
Here $\rho(r)$ is the
energy density, $p_r(r)$ is the radial pressure, and $p_t(r)$ is the transverse pressure.
Given the spherical symmetry,  $p_t(r)$ is the pressure 
measured in the two directions orthogonal to the radial direction. 
The  subscripts $\pm$ (on $m_\pm$ and $a_\pm$) have 
been (temporarily) suppressed for clarity.

The null energy condition (NEC) is satisfied provided, for any arbitrary null vector $k^a$, the stress-energy  $T_{\mu\nu}$ satisfies $T_{\mu\nu}\,k^\mu\,k^\nu\geq 0$. 
The radial null vector is $k^{\hat{\mu}}=(1,\pm 1,0,0)$ in the orthonormal frame where the stress energy is
$T_{\hat{\mu}\hat{\nu}}={\rm diag}[\rho(r),p_r(r),p_t(r),p_t(r)]$. Then
\begin{equation}
 (T_{\hat{\mu}\hat{\nu}}\,k^{\hat{\mu}}\,k^{\hat{\nu}})_\mathrm{radial}=\rho(r)+p_r(r)
 =-\frac{a^2(r-2m)}{4\pi r^{5}} < 0.    \label{generalNEC}
\end{equation}
To verify the negativity of this quantity, note that the bulk spacetime models a regular black hole when $a\in(0, 2m)$, with horizons at $u_{H} = \pm\sqrt{(2m)^{2}-a^2}$, corresponding to $r_H= u_H^2+a^2 = 2m$. We therefore `chop' the spacetime outside any horizons that are present. Hence in both of the bulk regions we have the condition that $r>2m$. 
Thus the radial NEC will be manifestly violated in both of the bulk regions of the black-bounce  thin-shell spacetime. In the context of static spherical symmetry, this is sufficient to conclude that all of the standard energy conditions associated with general relativistic analysis will be similarly violated.

In the transverse directions we can choose the null vector to be $k^{\hat{\mu}}=(1, 0, \cos\zeta,\sin\zeta)$ and so 
\begin{equation}
 (T_{\hat{\mu}\hat{\nu}}\,k^{\hat{\mu}}\,k^{\hat{\nu}})_\mathrm{transverse}
 =\rho(r)+p_t(r) =\frac{3 a^2m}{8\pi r^{5}}>0.    \label{generalNEC2}
\end{equation}
While this is manifestly positive this is not enough to override the NEC violations coming from the radial direction.

%###########################################################
\subsection{Normal 4-vector and extrinsic curvature}\label{SS:extrinsic}
%###########################################################

The two bulk manifolds,  ${\cal M_+}$ and ${\cal M_-}$, are bounded by the hypersurfaces $\partial{\cal M_+}=\Sigma_+$ and ${\cal M_-}=\Sigma_-$. 
These two hypersurfaces possess induced 3-metrics $g_{ij}^+$ and $g_{ij}^-$, respectively. The $\Sigma_{\pm}$ are chosen to be
isometric. (In terms of the intrinsic coordinates, $g_{ij}^{\pm}(\xi)=g_{ij}(\xi)$, with $\xi^i=(\tau,\theta,\phi)$.)  Thence a single 
manifold ${\cal M}$ is obtained by gluing together ${\cal M_+}$ and ${\cal M_-}$ at their 
boundaries, that is ${\cal M}={\cal M_+}\cup {\cal M_-}$, with the natural identification 
of the two boundaries $\Sigma_{\pm}=\Sigma$.

The boundary manifold $\Sigma$ possesses three tangent basis vectors ${\bf e}_{(i)}=\partial /\partial \xi^i$, with the holonomic components $e^{\mu}_{(i)}|_{\pm}=\partial x_{\pm}^{\mu}/\partial \xi^i$. This basis specifies the induced metric via the scalar product $g_{ij}={\bf e}_{(i)}\cdot {\bf e}_{(j)}=g_{\mu \nu}e^{\mu}_{(i)}e^{\nu}_{(j)}|_{\pm}$. Finally, in explicit coordinates, the intrinsic metric to $\Sigma$ is given by
\begin{equation}
ds^2_{\Sigma}=-d\tau^2 + R^2(\tau) \,(d\theta ^2+\sin
^2{\theta}\,d\phi^2).
\end{equation}
That is, the manifold ${\cal M}$ is obtained by gluing ${\cal M_+}$ and ${\cal M_-}$  
at the 3-surface $x^{\mu}(\tau,\theta,\phi)=(t(\tau),R(\tau),\theta,\phi)$. The respective $4$-velocities, tangent to the junction surface and orthogonal to the slices of spherical symmetry, are defined on the two sides of the junction surface. They are explicitly given by
\begin{eqnarray}
U^{\mu}_{\pm}=
\left(\frac{\sqrt{\left(1-\frac{2m_{\pm}}{R}\right)\left(1-\frac{a_{\pm}^2}{R^2}\right)+\dot{R}^{2}}}{\left(1-\frac{2m_{\pm}}{R}\right)\left(1-\frac{a_{\pm}^2}{R^2}\right)^{1/2}},\;
\dot{R},0,0 \right).
\end{eqnarray}
Here $\tau$ is the proper time of an
observer comoving with $\Sigma$, 
and the overdot denotes a derivative with respect to this proper time.
Furthermore, the timelike junction surface $\Sigma$ is given by the parametric equation $f(x^{\mu}(\xi^i))= r-a(\tau) = 0$, and the unit normal $4$-vector, $n^{\mu}$, is defined as
\begin{equation}\label{defnormal}
n_{\mu}=
{\nabla_a f\over ||\nabla f||} = 
\pm \,\left |g^{\alpha \beta}\,\frac{\partial f}{\partial
x ^{\alpha}} \, \frac{\partial f}{\partial x ^{\beta}}\right
|^{-1/2}\;\frac{\partial f}{\partial x^{\mu}}\,.
\end{equation}
Hence $n_{\mu}\,n^{\mu}=+1$ and $n_{\mu}e^{\mu}_{(i)}=0$. In the usual Israel formalism one chooses the
normals to point from ${\cal M_-}$ to ${\cal M_+}$ \cite{Israel}, so that the unit normals to the
junction surface are provided by the following expressions:
\begin{eqnarray}
n^{\mu}_{\pm}= \pm \left(
\frac{\dot{R}}{\left(1-\frac{2m_{\pm}}{R}\right)\left(1-\frac{a_{\pm}^2}{R^2}\right)^{1/2}},
\sqrt{\left(1-\frac{2m_{\pm}}{R}\right)\left(1-\frac{a_{\pm}^2}{R^2}\right)+\dot{R}^{2}},0,0
\right) \label{normal}
\,.
\end{eqnarray}
Taking into account spherical symmetry one may also obtain the above expressions from consideration of the contractions $U^{\mu}n_{\mu}=0$ and $n^{\mu}n_{\mu}=+1$.
The extrinsic curvature, or  second fundamental form, is typically defined as 
$K_{ij}=n_{{(}\mu;\nu{)}}e^{\mu}_{(i)}e^{\nu}_{(j)}$. 
Now, by differentiating $n_{\mu}e^{\mu}_{(i)}=0$ with respect to $\xi^j$,  one obtains the following useful relation
\begin{equation}
n_{\mu}\frac{\partial ^2 x^{\mu}}{\partial \xi^i \, \partial \xi^j}=
-n_{\mu,\nu}\, \frac{\partial x^{\mu}}{\partial \xi^i}\frac{\partial x^{\nu}}{\partial \xi^j},
\end{equation}
so that the extrinsic curvature  $K_{ij}$ can therefore be represented in the form
\begin{eqnarray}
\label{extrinsiccurv}
K_{ij}^{\pm}=-n_{\mu} \left(\frac{\partial ^2 x^{\mu}}{\partial
\xi ^{i}\,\partial \xi ^{j}}+\Gamma ^{\mu \pm}_{\;\;\alpha
\beta}\;\frac{\partial x^{\alpha}}{\partial \xi ^{i}} \,
\frac{\partial x^{\beta}}{\partial \xi ^{j}} \right) \,.
\end{eqnarray}
Finally, using both spherical symmetry and equation (\ref{extrinsiccurv}), the non-trivial components of the extrinsic curvature are given by:
\begin{eqnarray}
K^{\theta \;\pm }_{\;\;\theta}&=&\pm\frac{1}{R}\,\sqrt{\left(1-\frac{2m_{\pm}}{R}\right)\left(1-\frac{a_{\pm}^2}{R^2}\right)+\dot{R}^{2}}\;,
\label{genKplustheta}
\\
K^{\tau\;\pm}_{\;\;\tau}&=&\pm\,
\left[\frac{\ddot R-
\frac{a_{\pm}^2}{R\left(R^2 - a_{\pm}^2\right)} \dot{R^2}+
\frac{m_{\pm}\left(R^2 - a_{\pm}^2\right)}{R^4}}{\sqrt{\left(1-\frac{2m_{\pm}}{R}\right)\left(1-\frac{a_{\pm}^2}{R^2}\right)+\dot{R}^{2}}} \right]  \,.
\label{genKminustautau}
\end{eqnarray}

%###########################################################
\subsection{Lanczos equations and surface stress-energy}\label{SS:Lanczos}
%###########################################################
For the case of a thin shell, the extrinsic curvature need not be continuous across $\Sigma$. For notational clarity we denote the discontinuity in $K_{ij}$ as
$\kappa_{ij}=K_{ij}^{+}-K_{ij}^{-}$.
The Einstein equations, when applied to the hypersurface joining the bulk spacetimes, now yield the Lanczos equations:
\begin{equation}
S^{i}_{\;j}=-\frac{1}{8\pi}\,(\kappa ^{i}_{\;j}-\delta
^{i}_{\;j}\; \kappa ^{k}_{\;k})  \,,
\end{equation}
where $S^{i}_{\;j}$ is the surface stress-energy tensor on the junction interface $\Sigma$. Due to spherical symmetry $\kappa ^{i}_{\;j}={\rm diag} \left(\kappa ^{\tau}_{\;\tau},\kappa ^{\theta}_{\;\theta},\kappa^{\theta}_{\;\theta}\right)$, the surface stress-energy tensor reduces to $S^{i}_{\;j}={\rm diag}(-\sigma,{\cal P},{\cal P})$, where $\sigma$ is the surface energy density, and ${\cal P}$ the surface pressure.
The Lanczos equations imply
\begin{eqnarray}
\sigma &=&-\frac{1}{4\pi}\,\kappa ^{\theta}_{\;\theta} \,,\label{sigma} \\
{\cal P} &=&\frac{1}{8\pi}(\kappa ^{\tau}_{\;\tau}+\kappa
^{\theta}_{\;\theta}) \,. \label{surfacepressure}
\end{eqnarray}
Using the computed extrinsic curvatures (\ref{genKplustheta})--(\ref{genKminustautau}), 
we now evaluate the surface stresses:
\begin{eqnarray}
\sigma&=&-\frac{1}{4\pi R}\left[\sqrt{\left(1-\frac{2m_{+}}{R}\right)\left(1-\frac{a_{+}^2}{R^2}\right)+\dot{R}^{2}}+
\sqrt{\left(1-\frac{2m_{-}}{R}\right)\left(1-\frac{a_{-}^2}{R^2}\right)+\dot{R}^{2}}
\right],
\label{gen-surfenergy2}
\\
{\cal P}&=&\frac{1}{8\pi R}\left[
\frac{1+\dot{R}^2\left( \frac{R^2-2a_+^2}{R^2-a_+^2} \right)+R\ddot{R}-\frac{m_+ R^2 + a_+^2 \left(R-m_+\right)}{R^3}}{\sqrt{\left(1-\frac{2m_{+}}{R}\right)\left(1-\frac{a_{+}^2}{R^2}\right)+\dot{R}^{2}}}
+
\frac{1+\dot{R}^2\left( \frac{R^2-2a_-^2}{R^2-a_-^2} \right)+R\ddot{R}-\frac{m_- R^2 + a_-^2 \left(R-m_-\right)}{R^3}}{\sqrt{\left(1-\frac{2m_{-}}{R}\right)\left(1-\frac{a_{-}^2}{R^2}\right)+\dot{R}^{2}}}
\right].
\label{gen-surfpressure2}
\end{eqnarray}
The surface mass of the thin shell is defined by $m_s=4\pi R^2\sigma$, a result which we shall use extensively below.  Furthermore the surface energy density $\sigma$ is always negative, implying energy condition violations in this thin-shell context.
For the specific symmetric case, $m_+=m_-$, and for vanishing bounce parameters $a_{\pm}=0$, the analysis reduces to that of reference \citep{Poisson}. 

%#################################################################
\subsection{Gauss and Codazzi equations}\label{SS:Gauss}
%#################################################################

The Gauss equation is sometimes called the first contracted Gauss--Codazzi equation.
In standard general relativity it is more often referred to as the ``Hamiltonian constraint''.
The Gauss equation is a purely mathematical statement relating bulk curvature to extrinsic and intrinsic curvature at the boundary:
\begin{eqnarray}
G_{\mu \nu}\;n^{\mu}\,n^{\nu}=\frac{1}{2}\,(K^2-K_{ij}K^{ij}-\,^3R)\,.
    \label{1Gauss}
\end{eqnarray}
Applying the Einstein equation, and evaluating the discontinuity across the junction surface, this becomes
\begin{equation}
8\pi \left[T_{\mu \nu}n^{\mu}n^{\nu}\right]^{+}_{-}
 =
 {1\over2} [K^2-K_{ij}K^{ij}]\,.
\end{equation}
Using the conventions $\left[X \right]^+_-\equiv X^+|_{\Sigma}-X^-|_{\Sigma}$ and $\overline{X} \equiv \frac{1}{2}
(X^+|_{\Sigma}+X^-|_{\Sigma})$ for notational simplicity, and 
applying the Lanczos equations,  one deduces the constraint equation
\begin{eqnarray}
 \left[T_{\mu \nu}n^{\mu}n^{\nu} \right]^{+}_{-} = S^{ij}\;\overline{K}_{ij}\,.
\end{eqnarray}

In contrast the Codazzi equation (Codazzi--Mainardi equation), is often known as the 
second contracted Gauss--Codazzi equation.  In general relativity more often referred to as the ``ADM 
constraint'' or ``momentum constraint''.  The purely mathematical result is 
\begin{eqnarray}
G_{\mu \nu}\,e^{\mu}_{(i)}n^{\nu}=K^j_{i|j}-K,_{i}\,.
    \label{2Gauss}
\end{eqnarray}
Together with the Einstein and Lanczos equations, and considering the discontinuity across the thin shell, this now yields the conservation identity:
\begin{eqnarray}\label{conservation}
\left[T_{\mu \nu}\; e^{\mu}_{\;(j)}n^{\nu}\right]^+_- = - S^{i}_{\;j|i}\,.
\end{eqnarray}
The left-hand-side of the conservation identity (\ref{conservation}) can be interpreted in terms of 
 momentum flux.  Explicitly
\begin{eqnarray}
\left[T_{\mu\nu}\; e^{\mu}_{\;(\tau)}\,n^{\nu}\right]^+_-
=\left[T_{\mu\nu}\; U^{\mu}\,n^{\nu}\right]^+_-
&=&\left[\pm
\left(T_{\hat{t}\hat{t}}+T_{\hat{r}\hat{r}}\right)
\,\frac{\dot{R}\sqrt{\left(1-\frac{2m}{R}\right)\left( 1-\frac{a^2}{R^2} \right)+\dot{R}^{2}}}{\left(1-\frac{2m}{R}\right)\left( 1-\frac{a^2}{R^2} \right)} \;
\right]^+_-\,, 
	\nonumber  \\
	&=&\left[\mp
\frac{a^2}{4\pi R^4}
\,\frac{\dot{R}\sqrt{\left(1-\frac{2m}{R}\right)\left( 1-\frac{a^2}{R^2} \right)+\dot{R}^{2}}}{\left( 1-\frac{a^2}{R^2} \right)} \;
\right]^+_-\,, 
\label{flux}
\end{eqnarray}
where $T_{\hat{t}\hat{t}}$ and $T_{\hat{r}\hat{r}}$ are the bulk stress-energy tensor components
given in an orthonormal basis. Note that the flux term corresponds to the net discontinuity in the bulk
momentum flux $F_\mu=T_{\mu\nu}\,U^\nu$ which impinges on the shell. 
For notational simplicity, we write
\begin{equation}
\left[T_{\mu\nu}\; e^{\mu}_{\;(\tau)}\,n^{\nu}\right]^+_-
=\dot R\, \Xi\,.
\end{equation}
Here we have defined the useful quantity
\begin{equation}
\Xi
=- \frac{1}{4\pi R^2}
\left[\frac{a_+^2}{\left( R^2-a_+^2 \right)} \sqrt{\left(1-\frac{2m_+}{R}\right)\left( 1-\frac{a_+^2}{R^2} \right)+\dot{R}^{2}} + 
\frac{a_-^2}{\left( R^2-a_-^2 \right)} \sqrt{\left(1-\frac{2m_-}{R}\right)\left( 1-\frac{a_-^2}{R^2} \right)+\dot{R}^{2}} \;
\right] \,.
\end{equation}
Now $A=4\pi R^2$ is the surface area of the thin shell. 
The conservation identity becomes
\begin{equation}
\frac{d\sigma}{d\tau}+(\sigma+{\cal P})\,\frac{1}{A} \frac{dA}{d\tau}=\Xi \, \dot{R}\,.
\label{E:conservation2}
\end{equation}
Equivalently
\begin{equation}
\frac{d(\sigma A)}{d\tau}+{\cal P}\,\frac{dA}{d\tau}=\Xi \,A \, \dot{R}\,.
\label{E:conservation3}
\end{equation}
The term on the right-hand-side incorporates the flux term and encodes the work done by  external forces, while the first term on the left-hand-side is simply the variation of the internal energy of the shell, and the second term characterizes the work done by the shell's internal forces.
Provided that the equations of motion can be integrated to 
determine the surface energy density as a function of radius $R$ we infer the 
existence of a suitable function $\sigma(R)$.  Defining $ \sigma'=d\sigma /dR$ the conservation equation can then be written as
\begin{equation}
\sigma'=-\frac{2}{R}\,(\sigma +{\cal P})+\Xi\,.
\label{cons-equation}
\end{equation}

%##################################################################
\subsection{Equation of motion}\label{SS:eom}
%##################################################################

To analyze the stability of the wormhole, equation~(\ref{gen-surfenergy2}) can be rearranged to provide the  thin-shell equation of motion, given by
\begin{equation}
\frac{1}{2} \dot{R}^2+V(R)=0  \,.
\end{equation}
The potential $V(R)$ is defined as
\begin{equation}
V(R)= \frac{1}{2}\left\{ 1-{\bar \Delta(R)\over R} -\left[\frac{m_{s}(R)}{2R}\right]^2-\left[\frac{\Delta(R)}{m_{s}
(R)}\right]^2\right\}\,,
   \label{potential}
\end{equation}
where $m_s(R)=4\pi R^2\,\sigma(R)$ is the mass of the thin shell, and the quantities $\bar \Delta(R)$ and
$\Delta(R)$ are defined as
\begin{eqnarray}
\bar \Delta(R)&=&\left(m_{+}+m_{-}\right) + \frac{1}{2R} \left[ a_+^2 \left( 1-\frac{2m_+}{R} \right) + a_-^2 \left( 1-\frac{2m_-}{R} \right)  \right] ,\\
\Delta(R)&=&\left(m_{+} - m_{-}\right) + \frac{1}{2R} \left[ a_+^2 \left( 1-\frac{2m_+}{R} \right) - a_-^2 \left( 1-\frac{2m_-}{R} \right)  \right] ,
\end{eqnarray}
respectively.
Note that by differentiating with respect to {$\tau$}, the equation of motion implies $\ddot R = -V'({R}) $, which will be useful below.

As outlined in reference \cite{Montelongo-Garcia:2011}, we can reverse the logic flow and determine the surface mass as a function of the potential. More specifically, if we impose a specific potential $V({R})$, this potential implicitly tells us how much surface mass we need to distribute on the wormhole throat.
This further places implicit demands on the equation of state of the exotic matter residing on the wormhole throat. This implies that, after imposing the equation of motion for the shell, one has:

\medskip
\noindent
\emph{Surface energy density:}
\begin{equation}
\sigma=-\frac{1}{4\pi R}\left[\sqrt{\left(1-\frac{2m_{+}}{R}\right)\left(1-\frac{a_{+}^2}{R^2}\right) -2V(R)}+
\sqrt{\left(1-\frac{2m_{-}}{R}\right)\left(1-\frac{a_{-}^2}{R^2}\right)-2V(R)} \;
\right].
\label{gen-surfenergy2b}
\end{equation}
\emph{Surface pressure:}
\begin{eqnarray}
{\cal P}&=&\frac{1}{8\pi R}\left[
\frac{1-2V(R)\left( \frac{R^2-2a_+^2}{R^2-a_+^2} \right)-RV'(R)-\frac{m_+ R^2 + a_+^2 \left(R-m_+\right)}{R^3}}{\sqrt{\left(1-\frac{2m_{+}}{R}\right)
\left(1-\frac{a_{+}^2}{R^2}\right){-2V(R)}}}
	\right.
	\nonumber \\
&&\qquad  \qquad \left.
+
\frac{1-2V(R)\left( \frac{R^2-2a_-^2}{R^2-a_-^2} \right)-RV'(R)-\frac{m_- R^2 + a_-^2 \left(R-m_-\right)}{R^3}}{\sqrt{\left(1-\frac{2m_{-}}{R}\right)
\left(1-\frac{a_{-}^2}{R^2}\right){-2V(R)}}}
\right] \,.
\label{gen-surfpressure2b}
\end{eqnarray}
\emph{External energy flux:}
\begin{equation}
\Xi
=- \frac{1}{4\pi R^2}
\left[\frac{a_+^2}{\left( R^2-a_+^2 \right)} \sqrt{\left(1-\frac{2m_+}{R}\right)\left( 1-\frac{a_+^2}{R^2} \right)
-2V(R)} + 
\frac{a_-^2}{\left( R^2-a_-^2 \right)} \sqrt{\left(1-\frac{2m_-}{R}\right)\left( 1-\frac{a_-^2}{R^2} \right)
-2V(R)} \;
\right] \,.
\end{equation}
These three quantities, $\{\sigma({R}),\,{\cal P}({R}),\,\Xi({R})\}$, are inter-related by the differential conservation law, so at most two of them are functionally independent. We could equivalently work with the quantities $\{m_s({R}),\,{\cal P}({R}),\,\Xi({R})\}$.

%###################################################################
\subsection{Linearized equation of motion}\label{SS:linearized}
%###################################################################

We now consider the equation of motion $\frac{1}{2}\dot R^2 + V(R)=0$, which implies $\ddot R = -V'(R)$,
and linearize around an assumed static solution at $R_0$.
This implies that a second-order Taylor expansion of $V(R)$ around $R_0$ provides
\begin{equation}
V(R)=V(R_0)+V'(R_0)(R-R_0)+\frac{1}{2}V''(R_0)(R-R_0)^2+O[(R-R_0)^3]
\,.   \label{linear-potential0}
\end{equation}
Since we are expanding around a static solution, $\dot R_0=\ddot R_0 = 0$, we have both $V(R_0)=V'(R_0)=0$, so that equation (\ref{linear-potential0}) reduces to
\begin{equation}
V(R)= \frac{1}{2}V''(R_0)(R-R_0)^2+O[(R-R_0)^3]
\,.   \label{linear-potential}
\end{equation}
The static solution at $R_0$ is stable if and only if $V(R)$ has a local minimum at $R_0$.
This requires $V''(R_{0})>0$. This stability condition will be our fundamental tool in the subsequent analysis --- though reformulation in terms of more basic quantities will prove useful.
For instance, it is useful to express the quantities $m_s'(R)$ and $m_s''(R)$ in terms of the potential and its derivatives ---  doing so allows us  to develop a simple inequality on
$m_s''(R_0)$ by using the constraint $V''(R_0)>0$.
Similar formulae will hold for the pairs $\sigma'(R)$, $\sigma''(R)$, for ${\cal P}'(R)$, ${\cal P}''(R)$, and for $\Xi'(R)$, $\Xi''(R)$. In view of the multiple redundancies coming from the relations $m_s(R) = 4\pi\sigma(R) R^2$ and the differential conservation law, we can easily see that the only interesting quantities are  $\Xi'(R)$, $\Xi''(R)$.

In the applications analysed below, it is extremely useful to  consider the dimensionless quantity
\begin{equation}
{m_s(R)\over R} =4\pi \sigma(R) R = - \left[\sqrt{\left(1-\frac{2m_{+}}{R}\right)\left(1-\frac{a_{+}^2}{R^2}\right) -2V(R)}+
\sqrt{\left(1-\frac{2m_{-}}{R}\right)\left(1-\frac{a_{-}^2}{R^2}\right)-2V(R)} \;
\right].
\end{equation}
We now express $[m_s(R)/R]'$ and $[m_s(R)/R]''$ in terms of the following quantities:
\begin{equation}
\left[m_s(R)\over R\right]' = -  \left\{
{ \frac{m_+}{R^2}\left(1-\frac{a_{+}^2}{R^2}\right) + \frac{a_+^2}{R^3}\left(1-\frac{2m_{+}}{R}\right)- V'(R)
\over
\sqrt{\left(1-\frac{2m_{+}}{R}\right)\left(1-\frac{a_{+}^2}{R^2}\right)-2V(R)}} 
+
{ \frac{m_-}{R^2}\left(1-\frac{a_{-}^2}{R^2}\right) + \frac{a_-^2}{R^3}\left(1-\frac{2m_{-}}{R}\right)- V'(R)
\over
\sqrt{\left(1-\frac{2m_{-}}{R}\right)\left(1-\frac{a_{-}^2}{R^2}\right)-2V(R)}}  \right\} \,,
\end{equation}
and
\begin{eqnarray}
\left[\frac{m_s(R)}{R}\right]'' =   \left\{
{\left[ \frac{m_+}{R^2}\left(1-\frac{a_{+}^2}{R^2}\right) + \frac{a_+^2}{R^3}\left(1-\frac{2m_{+}}{R}\right)- V'(a) \right]^2
\over
\left[\left(1-\frac{2m_{+}}{R}\right)\left(1-\frac{a_{+}^2}{R^2}\right)-2V(R)\right]^{3/2}}
+
{ \frac{2m_+}{R^3}\left(1-\frac{a_{+}^2}{R^2}\right) + \frac{3a_+^2}{R^4}\left(1-\frac{2m_{+}}{R}\right) - \frac{4m_+ a_+^2}{R^5} + V''(R)
\over
\sqrt{\left(1-\frac{2m_{+}}{R}\right)\left(1-\frac{a_{+}^2}{R^2}\right)-2V(R)}} 
   \right.
	\nonumber \\
	\left.
+
{\left[ \frac{m_-}{R^2}\left(1-\frac{a_{-}^2}{R^2}\right) + \frac{a_-^2}{R^3}\left(1-\frac{2m_{-}}{R}\right)- V'(R) \right]^2
\over
\left[\left(1-\frac{2m_{-}}{R}\right)\left(1-\frac{a_{-}^2}{R^2}\right)-2V(R)\right]^{3/2}}
+
{ \frac{2m_-}{R^3}\left(1-\frac{a_{-}^2}{R^2}\right) + \frac{3a_-^2}{R^4}\left(1-\frac{2m_{-}}{R}\right) - \frac{4m_{-} a_{-}^2}{R^5} + V''(R)
\over
\sqrt{\left(1-\frac{2m_{-}}{R}\right)\left(1-\frac{a_{-}^2}{R^2}\right)-2V(R)}}   \right\} \,.
\nonumber\\
&&
\end{eqnarray}
Similarly, consider the useful dimensionless quantity 
\begin{equation}
4\pi R^2\, \Xi
=- 
\left[\frac{a_+^2}{\left( R^2-a_+^2 \right)} \sqrt{\left(1-\frac{2m_+}{R}\right)\left( 1-\frac{a_+^2}{R^2} \right)
-2V(R)} + 
\frac{a_-^2}{\left( R^2-a_-^2 \right)} \sqrt{\left(1-\frac{2m_-}{R}\right)\left( 1-\frac{a_-^2}{R^2} \right)
-2V(R)} \;
\right] \,.
\end{equation}
This leads to the following relations:
\begin{eqnarray}
&\left[4\pi R^2\, \Xi \right]'
= \frac{a_+^2}{\left( R^2-a_+^2 \right)}
\left[\frac{2R}{\left( R^2-a_+^2 \right)} \sqrt{\left(1-\frac{2m_+}{R}\right)\left( 1-\frac{a_+^2}{R^2} \right)
-2V(R)} -
{ \frac{m_+}{R^2}\left(1-\frac{a_{+}^2}{R^2}\right) + \frac{a_+^2}{R^3}\left(1-\frac{2m_{+}}{R}\right)- V'(R)
\over
\sqrt{\left(1-\frac{2m_{+}}{R}\right)\left(1-\frac{a_{+}^2}{R^2}\right)-2V(R)}}  \right]
\nonumber \\
&+\frac{a_-^2}{\left( R^2-a_-^2 \right)}
\left[\frac{2R}{\left( R^2-a_-^2 \right)} \sqrt{\left(1-\frac{2m_-}{R}\right)\left( 1-\frac{a_-^2}{R^2} \right)
-2V(R)} -
{ \frac{m_-}{R^2}\left(1-\frac{a_{-}^2}{R^2}\right) + \frac{a_-^2}{R^3}\left(1-\frac{2m_{-}}{R}\right)- V'(R)
\over
\sqrt{\left(1-\frac{2m_{-}}{R}\right)\left(1-\frac{a_{-}^2}{R^2}\right)-2V(R)}}  \right] \,,
\end{eqnarray}
and
\begin{eqnarray}
&\left[4\pi R^2\, \Xi \right]''
= 
\left\{-\frac{{2}a_{+}^2 ({3}R^2+a_{+}^2)}{\left( R^2-a_{+}^2 \right)^3} \sqrt{\left(1-\frac{2m_+}{R}\right)\left( 1-\frac{a_+^2}{R^2} \right)-2V(R)} 
+
\frac{4a_{+}^2 R}{\left( R^2-a_{+}^2 \right)^2} 
{ \frac{m_+}{R^2}\left(1-\frac{a_{+}^2}{R^2}\right) + \frac{a_+^2}{R^3}\left(1-\frac{2m_{+}}{R}\right)- V'(R)
\over
\sqrt{\left(1-\frac{2m_{+}}{R}\right)\left(1-\frac{a_{+}^2}{R^2}\right)-2V(R)}}  
	\right.
	\nonumber \\
&  \left. +
\frac{a_{+}^2 }{\left( R^2-a_{+}^2 \right)} 
{\left[ \frac{m_+}{R^2}\left(1-\frac{a_{+}^2}{R^2}\right) + \frac{a_+^2}{R^3}\left(1-\frac{2m_{+}}{R}\right)- V'(R)\right]^2
\over
\left[\left(1-\frac{2m_{+}}{R}\right)\left(1-\frac{a_{+}^2}{R^2}\right)-2V(R)\right]^{3(2}}  
+
\frac{a_{+}^2 }{\left( R^2-a_{+}^2 \right)} 
{ \frac{2m_+}{R^3}\left(1-\frac{a_{+}^2}{R^2}\right) + \frac{3a_+^2}{R^4}\left(1-\frac{2m_{+}}{R}\right) - \frac{4m_+ a_+^2}{R^5} + V''({R})
\over
\sqrt{\left(1-\frac{2m_{+}}{R}\right)\left(1-\frac{a_{+}^2}{R^2}\right)-2V(R)}}  \right\}
\nonumber  \\
&	
%%%%
+
	\left\{-\frac{{2}a_{-}^2 ({3}R^2+a_{-}^2)}{\left( R^2-a_{-}^2 \right)^3} \sqrt{\left(1-\frac{2m_-}{R}\right)\left( 1-\frac{a_-^2}{R^2} \right)-2V(R)} 
+
\frac{4a_{-}^2 R}{\left( R^2-a_{-}^2 \right)^2} 
{ \frac{m_-}{R^2}\left(1-\frac{a_{-}^2}{R^2}\right) + \frac{a_-^2}{R^3}\left(1-\frac{2m_{-}}{R}\right)- V'(R)
\over
\sqrt{\left(1-\frac{2m_{-}}{R}\right)\left(1-\frac{a_{-}^2}{R^2}\right)-2V(R)}}  
	\right.
	\nonumber \\
&  \left. +
\frac{a_{-}^2 }{\left( R^2-a_{-}^2 \right)} 
{\left[ \frac{m_-}{R^2}\left(1-\frac{a_{-}^2}{R^2}\right) + \frac{a_-^2}{R^3}\left(1-\frac{2m_{-}}{R}\right)- V'(R)\right]^2
\over
\left[\left(1-\frac{2m_{-}}{R}\right)\left(1-\frac{a_{-}^2}{R^2}\right)-2V(R)\right]^{3(2}}  
+
\frac{a_{-}^2 }{\left( R^2-a_{-}^2 \right)} 
{ \frac{2m_-}{R^3}\left(1-\frac{a_{-}^2}{R^2}\right) + \frac{3a_-^2}{R^4}\left(1-\frac{2m_{-}}{R}\right) - \frac{4m_- a_-^2}{R^5} + V''(R)
\over
\sqrt{\left(1-\frac{2m_{-}}{R}\right)\left(1-\frac{a_{-}^2}{R^2}\right)-2V(R)}}  \right\}\,.
\end{eqnarray}

%######################################################################
\subsection{Master equation}\label{SS:master}
%%#####################################################################

Taking into account the extensive discussion above, we see that to have a stable static solution at ${R}_0$, we must satisfy two equations and one inequality. Specifically:
\begin{equation}
{m_s(R_0)\over R_0}  =4\pi \sigma(R_0) R_0 = - \left[\sqrt{\left(1-\frac{2m_{+}}{R_0}\right)\left(1-\frac{a_{+}^2}{R_0^2}\right) }+
\sqrt{\left(1-\frac{2m_{-}}{R_0}\right)\left(1-\frac{a_{-}^2}{R_0^2}\right)} \;
\right],
\end{equation}
and
\begin{equation}
\left[\frac{m_s(R_0)}{R_0}\right]' = -  \left\{
{ \frac{m_+}{R_0^2}\left(1-\frac{a_{+}^2}{R_0^2}\right) + \frac{a_+^2}{R_0^3}\left(1-\frac{2m_{+}}{R_0}\right)
\over
\sqrt{\left(1-\frac{2m_{+}}{R_0}\right)\left(1-\frac{a_{+}^2}{R_0^2}\right)}} 
+
{ \frac{m_-}{R_0^2}\left(1-\frac{a_{-}^2}{R_0^2}\right) + \frac{a_-^2}{R_0^3}\left(1-\frac{2m_{-}}{R_0}\right)
\over
\sqrt{\left(1-\frac{2m_{-}}{R_0}\right)\left(1-\frac{a_{-}^2}{R_0^2}\right)}}  \right\} ,
\end{equation}
and
\begin{eqnarray}
\left[\frac{m_s(R_0)}{R_0}\right]''    \geq    \left\{
{\left[ \frac{m_+}{R_0^2}\left(1-\frac{a_{+}^2}{R_0^2}\right) + \frac{a_+^2}{R_0^3}\left(1-\frac{2m_{+}}{R_0}\right) \right]^2
\over
\left[\left(1-\frac{2m_{+}}{R_0}\right)\left(1-\frac{a_{+}^2}{R_0^2}\right)\right]^{3/2}}
+
{ \frac{2m_+}{R_0^3}\left(1-\frac{a_{+}^2}{R_0^2}\right) + \frac{3a_+^2}{R_0^4}\left(1-\frac{2m_{+}}{R_0}\right) - \frac{4m_+ a_+^2}{R_0^5} 
\over
\sqrt{\left(1-\frac{2m_{+}}{R_0}\right)\left(1-\frac{a_{+}^2}{R_0^2}\right)}} 
   \right.
	\nonumber \\
	\left.
+
{\left[ \frac{m_-}{R_0^2}\left(1-\frac{a_{-}^2}{R_0^2}\right) + \frac{a_-^2}{R_0^3}\left(1-\frac{2m_{-}}{R_0}\right) \right]^2
\over
\left[\left(1-\frac{2m_{-}}{R_0}\right)\left(1-\frac{a_{-}^2}{R_0^2}\right)\right]^{3/2}}
+
{ \frac{2m_-}{R_0^3}\left(1-\frac{a_{-}^2}{R_0^2}\right) + \frac{3a_-^2}{R_0^4}\left(1-\frac{2m_{-}}{R_0}\right) - \frac{4m_{-} a_{-}^2}{R_0^5} 
\over
\sqrt{\left(1-\frac{2m_{-}}{R_0}\right)\left(1-\frac{a_{-}^2}{R_0^2}\right)}}   \right\} .
\label{stability_constraint1}
\end{eqnarray}
More specifically, this last inequality translates the stability condition $V''(R_0) \geq 0$ into an explicit inequality on $m_s''({R}_0)$, an inequality that can in particular cases be explicitly checked.
In the absence of external forces this inequality is the only stability condution one requires. 
However, once one has external forces (that is, in the presence of fluxes $\Xi \neq 0$),  there is additional information:
\begin{equation}
4\pi R_0^2\, \Xi_0
=- 
\left[\frac{a_+^2}{\left( R_0^2-a_+^2 \right)} \sqrt{\left(1-\frac{2m_+}{R_0}\right)\left( 1-\frac{a_+^2}{R_0^2} \right)} + 
\frac{a_-^2}{\left( R_0^2-a_-^2 \right)} \sqrt{\left(1-\frac{2m_-}{R_0}\right)\left( 1-\frac{a_-^2}{R_0^2} \right)} \;
\right] \,.
\end{equation}
This leads one to consider the quantity
\begin{eqnarray}
\left[4\pi R_0^2\, \Xi_0 \right]'
&=& \frac{a_+^2}{\left( R_0^2-a_+^2 \right)}
\left[\frac{2R_0}{\left( R_0^2-a_+^2 \right)} \sqrt{\left(1-\frac{2m_+}{R_0}\right)\left( 1-\frac{a_+^2}{R_0^2} \right)} -
{ \frac{m_+}{R_0^2}\left(1-\frac{a_{+}^2}{R_0^2}\right) + \frac{a_+^2}{R_0^3}\left(1-\frac{2m_{+}}{R_0}\right)
\over
\sqrt{\left(1-\frac{2m_{+}}{R_0}\right)\left(1-\frac{a_{+}^2}{R_0^2}\right)}}  \right]
%
%\right.
\nonumber \\
%\left.
&&+\frac{a_-^2}{\left( R_0^2-a_-^2 \right)}
\left[\frac{2R_0}{\left( R_0^2-a_-^2 \right)} \sqrt{\left(1-\frac{2m_-}{R_0}\right)\left( 1-\frac{a_-^2}{R_0^2} \right)} -
{ \frac{m_-}{R_0^2}\left(1-\frac{a_{-}^2}{R_0^2}\right) + \frac{a_-^2}{R_0^3}\left(1-\frac{2m_{-}}{R_0}\right)
\over
\sqrt{\left(1-\frac{2m_{-}}{R_0}\right)\left(1-\frac{a_{-}^2}{R_0^2}\right)}}  \right] \,.
\end{eqnarray}
Furthermore, note that since $R_{0} > a_{\pm}$, the inequality on $\left[4\pi R_0^2\, \Xi_0 \right]''$ is given by
\begin{eqnarray}
\left[4\pi R_0^2\, \Xi_0 \right]''
\geq 
\left\{-\frac{3a_{+}^2 (2R_0^2+a_{+}^2)}{\left( R_0^2-a_{+}^2 \right)^3} \sqrt{\left(1-\frac{2m_+}{R_0}\right)\left( 1-\frac{a_+^2}{R_0^2} \right)} 
+
\frac{4a_{+}^2 R_0}{\left( R_0^2-a_{+}^2 \right)^2} 
{ \frac{m_+}{R_0^2}\left(1-\frac{a_{+}^2}{R_0^2}\right) + \frac{a_+^2}{R_0^3}\left(1-\frac{2m_{+}}{R_0}\right)
\over
\sqrt{\left(1-\frac{2m_{+}}{R_0}\right)\left(1-\frac{a_{+}^2}{R_0^2}\right)}}  
	\right.
	\nonumber \\
  \left. +
\frac{a_{+}^2 }{\left( R_0^2-a_{+}^2 \right)} 
{\left[ \frac{m_+}{R_0^2}\left(1-\frac{a_{+}^2}{R_0^2}\right) + \frac{a_+^2}{R_0^3}\left(1-\frac{2m_{+}}{R_0}\right)\right]^2
\over
\left[\left(1-\frac{2m_{+}}{R_0}\right)\left(1-\frac{a_{+}^2}{R_0^2}\right)\right]^{3/2}}  
+
\frac{a_{+}^2 }{\left( R_0^2-a_{+}^2 \right)} 
{ \frac{2m_+}{R_0^3}\left(1-\frac{a_{+}^2}{R_0^2}\right) + \frac{3a_+^2}{R_0^4}\left(1-\frac{2m_{+}}{R_0}\right) - \frac{4m_+ a_+^2}{R_0^5}
\over
\sqrt{\left(1-\frac{2m_{+}}{R_0}\right)\left(1-\frac{a_{+}^2}{R_0^2}\right)}}  \right\}
\nonumber  \\	
%%%%
+
	\left\{-\frac{3a_{-}^2 (2R_0^2+a_{-}^2)}{\left( R_0^2-a_{-}^2 \right)^3} \sqrt{\left(1-\frac{2m_-}{R_0}\right)\left( 1-\frac{a_-^2}{R_0^2} \right)} 
+
\frac{4a_{-}^2 R_0}{\left( R_0^2-a_{-}^2 \right)^2} 
{ \frac{m_-}{R_0^2}\left(1-\frac{a_{-}^2}{R_0^2}\right) + \frac{a_-^2}{R_0^3}\left(1-\frac{2m_{-}}{R_0}\right)
\over
\sqrt{\left(1-\frac{2m_{-}}{R_0}\right)\left(1-\frac{a_{-}^2}{R_0^2}\right)}}  
	\right.
	\nonumber \\
  \left. +
\frac{a_{-}^2 }{\left( R_0^2-a_{-}^2 \right)} 
{\left[ \frac{m_-}{R_0^2}\left(1-\frac{a_{-}^2}{R_0^2}\right) + \frac{a_-^2}{R_0^3}\left(1-\frac{2m_{-}}{R_0}\right)\right]^2
\over
\left[\left(1-\frac{2m_{-}}{R_0}\right)\left(1-\frac{a_{-}^2}{R_0^2}\right)\right]^{3/2}}  
+
\frac{a_{-}^2 }{\left( R_0^2-a_{-}^2 \right)} 
{ \frac{2m_-}{R_0^3}\left(1-\frac{a_{-}^2}{R_0^2}\right) + \frac{3a_-^2}{R_0^4}\left(1-\frac{2m_{-}}{R_0}\right) - \frac{4m_- a_-^2}{R_0^5} 
\over
\sqrt{\left(1-\frac{2m_{-}}{R_0}\right)\left(1-\frac{a_{-}^2}{R_0^2}\right)}}  \right\}\,.
	\label{stability_constraint2}
\end{eqnarray}
In summary, the inequalities (\ref{stability_constraint1}) and (\ref{stability_constraint2}) dictate the stability regions of the wormhole solutions considered in this work, and in the following section we consider specific applications and examples.

%=================================================================
\section{Applications and Examples}\label{S:applications}
%=================================================================

In this section, we shall apply the general formalism described above to some specific examples. Several of these special cases are particularly important in order to emphasize the specific features of these black bounce spacetimes. Some examples are essential to assess the simplifications due to symmetry between the two asymptotic regions, while other cases are useful to understand the asymmetry between the two universes used in traversable wormhole construction.
In the following analysis we will consider specific cases by tuning the parameters of the bulk spacetimes, namely, the bounce parameters $a_{\pm}$, and the masses $m_{\pm}$.

%####################################################################
\subsection{Vanishing flux term: $a_{\pm}=0$}\label{SS1a:specific}
%####################################################################

\begin{figure}[!h]
\includegraphics[scale=0.470]{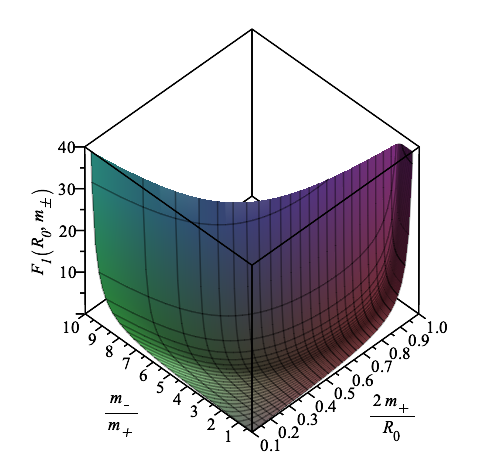}
\includegraphics[scale=0.470]{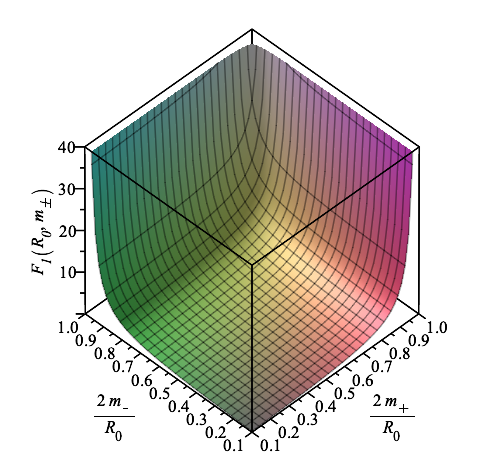}
\caption{Stability analysis for thin-shell Schwarzschild traversable wormholes, taking into account that $a_{\pm}=0$ and $m_{-} \neq  m_{+}$. 
The surfaces are given by the dimensionless quantity $F_{1}(R_0,m_\pm)	$, 	
defined by the right-hand-side of inequality (\ref{stabilitySchw}). 
The stability regions lie above the surfaces depicted  in the plots. 
We have considered the range $0<x=2m_+/R_0 <1$ and $0< y_1=m_-/m_+ < 1/x$, 
in the left plot and  $0<y_2=2m_-/R_0 <1$ in the right plot, respectively. 
Note that large stability regions exist for low values of $x=2m_+/R_{0}$ and of $y_{1,2}$. 
For regions close to the event horizon, $x \rightarrow 1$, the stability region decreases in size and only increases significantly for low values of $y_1$. See the text for details.}
\label{fig:stable1}
\end{figure}

Here, we consider the case of a vanishing flux term, that is $\Xi =0$, which is induced by imposing $a_{\pm}=0$. Thus, the only stability constraint arises from inequality (\ref{stability_constraint1}). Note that this case corresponds to the thin-shell Schwarzschild traversable wormholes analysed in references~\cite{Poisson,Montelongo-Garcia:2011}.
For the specific case of $a_{\pm}=0$, and considering an asymmetry in the masses $m_{-} \neq  m_{+}$, inequality (\ref{stability_constraint1}) reduces to 
\begin{equation}
R_0^2 \left[\frac{m_s(R_0)}{R_0}\right]''    \geq  F_{1}(R_0,m_{\pm}) = \frac{\frac{2m_{+}}{R_0} \left( 1-\frac{3m_{+}}{2R_0} \right)}{\left( 1- \frac{2m_{+}}{R_0} \right)^{3/2}} +
\frac{\frac{2m_{-}}{R_0} \left( 1-\frac{3m_{-}}{2R_0} \right)}{\left( 1- \frac{2m_{-}}{R_0} \right)^{3/2}} \,.
\label{stabilitySchw}
\end{equation}
Note that in order to plot the stability regions, we have defined the following dimensionless form of the constraint as $F_{1}(R_0,m_{\pm})= R_0^2 \left[m_s(R_0)/R_0\right]''  $, which is depicted as the surfaces given in the plots of figure~\ref{fig:stable1}. The stability regions lie above these surfaces. In order to visualize the whole range of the parameters, so as to bring infinite $R_0$ in to a finite region of the plot, we have considered the definition $x=2m_{+}/R_0$ for convenience. For instance, the limit  $R_0 \rightarrow \infty$ corresponds to $x\rightarrow 0$, and $R_0 = 2m_{+}$ is equivalent to $x=1$. Thus, we have considered the range $0<x <1$. 

In the left plot of figure~\ref{fig:stable1}, we have considered the parameter $y_1=m_-/m_+$, which lies within the range $0< y_1 < 1/x$.  This parameter provides information on the relative variation of the masses. However, one may also consider a more symmetrical form of the stability analysis, by considering the definition  $y_2=2m_{-}/R_0$, which possesses the range $0<y_2 <1$, and the stability region is depicted in the right plot of figure~\ref{fig:stable1}. These two plots provide complementary information.

Regarding  the stability of the solution, from figure~\ref{fig:stable1} we verify that large stability regions exist for low values of $x=2m_+/R_0$ and of $y_1=m_-/m_+$ (and of $y_2=2m_-/R_0$). For regions close to the event horizon, $x \rightarrow 1$, the stability region decreases in size and only exists for low values of $y_{1,2}$. The specific case of $y_1=m_-/m_+=1$, corresponds to the thin-shell Schwarzschild wormholes analysed in \cite{Montelongo-Garcia:2011}, and one verifies that the size of the stability regions increases as the junction interface of the thin-shell increases. Namely, as $x=2m_+/R_0 \rightarrow 0$, as is transparent from figure~\ref{fig:stable1}.
%=============================================================
%=============================================================

%####################################################################
\subsection{Vanishing mass: $a_{+} \neq a_{-}$ and $m_{\pm}=0$}\label{SS2a:symmetric}
%####################################################################

Consider now the case of vanishing mass terms $m_{\pm}=0$, with an asymmetry of the bounce parameters $a_{+} \neq a_{-}$.

For this case, inequality (\ref{stability_constraint1}) reduces to 
\begin{equation}
R_0^2 \left[\frac{m_s(R_0)}{R_0}\right]''    \geq  F_{2}(R_0,a_{\pm}) = 2 \left[ \frac{\frac{a_{+}^2}{R_0^2} \left( \frac{3}{2} -\frac{a_{+}^2}{R_0^2} \right)}{\left( 1- \frac{a_{+}^2}{R_0^2} \right)^{3/2}} +
\frac{\frac{a_{-}^2}{R_0^2} \left( \frac{3}{2} -\frac{a_{-}^2}{R_0^2} \right)}{\left( 1- \frac{a_{-}^2}{R_0^2} \right)^{3/2}}  \right] \,,
\label{stabilityzeromass1}
\end{equation}
and inequality (\ref{stability_constraint2}) takes the following form
\begin{equation}
R_0^2 \left[4\pi R_0^2\, \Xi(R_0)\right]''    \geq  G_{2}(R_0,a_{\pm}) = - \frac{\frac{a_{+}^2}{R_0^2} \left( 6 -4\frac{a_{+}^2}{R_0^2} + 2 \frac{a_{+}^4}{R_0^4} \right)}{\left( 1- \frac{a_{+}^2}{R_0^2} \right)^{5/2}} 
-
\frac{\frac{a_{-}^2}{R_0^2} \left( 6 - 4\frac{a_{-}^2}{R_0^2} + 2 \frac{a_{-}^4}{R_0^4} \right)}{\left( 1- \frac{a_{-}^2}{R_0^2} \right)^{5/2}}   \,.
\label{stabilityzeromass2}
\end{equation}

\begin{figure}[!h]
\includegraphics[scale=0.50]{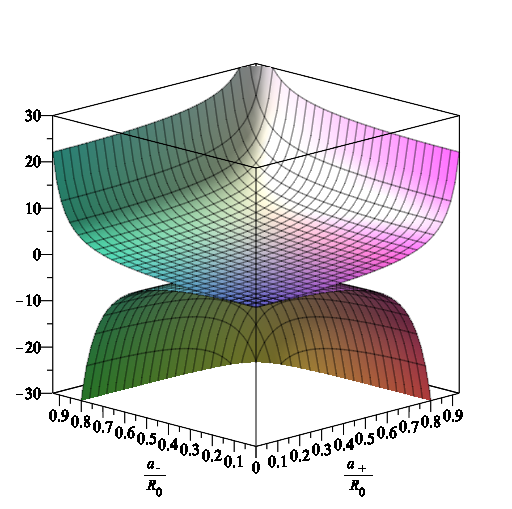}
\caption{The upper surface depicts the quantity $F_{2}(R_0,a_{\pm})= R_0^2 \left[m_s(R_0)/R_0\right]''  $, and the stable region lies above the surface of that curve. On the other hand, the function $G_{2}(R_0,a_{\pm})=R_0^2 \left[4\pi R_0^2\, \Xi_0 \right]''$ is depicted by the lower surface, and the stable region also lies above the surface of that curve. 
Thus the final stability region of the solution lies above the upper surface. 
See the text for more details.}
\label{fig:zeromass}
\end{figure}

We now consider the definition of the parameters $x=a_+/R_0$ and $y=a_-/R_0$ for convenience, so as to bring infinite $R_0$ within a finite region of the plot. That is,  $R_0 \rightarrow \infty$ is represented as $x\rightarrow 0$; and $R_0 = a_{+}, a_{-}$ is equivalent to $x, y=1$. Thus, the parameters $x$ and $y$ are restricted to the ranges $0<x<1$ and $0<y<1$.
Inequality (\ref{stabilityzeromass1}) is depicted as the upper surface in figure~\ref{fig:zeromass} and inequality (\ref{stabilityzeromass2}) depicted as the lower surface, and the stability regions are given above the respective surfaces. Thus, the final stability region lies above the upper surface.

%=============================================================
%=============================================================

%####################################################################
\subsection{Asymmetric vanishing parameters: $a_{+}=0$ and $m_{-}=0$}\label{SS2:asymmetric}
%####################################################################

Consider now the case of vanishing interior mass $m_{-}=0$, and vanishing exterior parameter $a_{+}=0$.
For this case, the inequality (\ref{stability_constraint1}) reduces to 
\begin{equation}
R_0^2 \left[\frac{m_s(R_0)}{R_0}\right]''    \geq  F_{3}(R_0,m_{+},a_{-}) =  \left[ \frac{\frac{2m_{+}}{R_0} \left( 1 -\frac{3m_{+}}{2R_0} \right)}{\left( 1- \frac{2m_{+}}{R_0} \right)^{3/2}} 
+
\frac{\frac{3a_{-}^2}{R_0^2} \left( 1 -\frac{2a_{-}^2}{3R_0^2} \right)}{\left( 1- \frac{a_{-}^2}{R_0^2} \right)^{3/2}}  \right] \,,
\label{stabilityzerom_zeroA1}
\end{equation}
and inequality (\ref{stability_constraint2}) is given by
\begin{equation}
R_0^2 \left[4\pi R_0^2\, \Xi(R_0)\right]''    \geq  G_{3}(R_0,a_{-}) = -
\frac{2\frac{a_{-}^2}{R_0^2} \left( 3 - 2\frac{a_{-}^2}{R_0^2} +  \frac{a_{-}^4}{R_0^4} \right)}{\left( 1- \frac{a_{-}^2}{R_0^2} \right)^{5/2}}   \,.
\label{stabilityzerom_zeroA2}
\end{equation}

\begin{figure}[!h]
\includegraphics[scale=0.50]{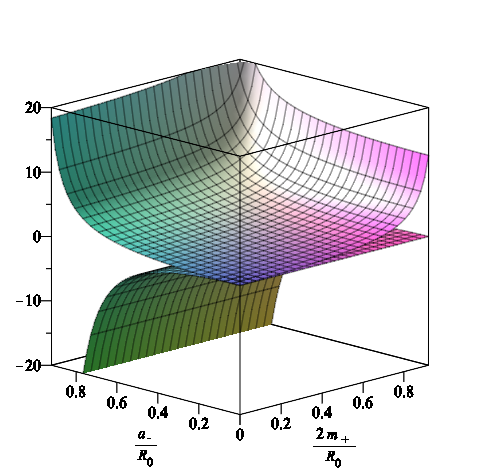}
\caption{The upper surface depicts the quantity $F_{3}(R_0,m_{+},a_{-})= R_0^2 \left[m_s(R_0)/R_0\right]''  $, and the stable region lies above that surface. 
On the other hand, the function $G_{3}(R_0,a_{-})=R_0^2 \left[4\pi R_0^2\, \Xi_0 \right]''$ is depicted by the lower surface, and the stable region is also lies above that surface of that plot. Thus, the final stability region of the solution lies above the upper surface. 
See the text for more details.}
\label{fig_zerom_zeroA.png}
\end{figure}

These are depicted as the upper and lower surfaces, respectively, in figure~\ref{fig_zerom_zeroA.png}. As in the previous example, the final stability region of the solution lies above the upper surface.

%=============================================================
%=============================================================

%####################################################################
\subsection{Mirror symmetry: $a_{\pm}=a$ and $m_{\pm}=m$}\label{SS3a:symmetric}
%####################################################################
Consider, for simplicity, the symmetric case, i.e., $a_{\pm}=a$ and $m_{\pm}=m$, so that the stability conditions reduce to:
\begin{eqnarray}
R_0^2 \left[\frac{m_s(R_0)}{R_0}\right]''    
\geq  
2   \left\{
{\left[ \frac{m}{R_0}\left(1-\frac{a^2}{R_0^2}\right) + \frac{a^2}{R_0^2}\left(1-\frac{2m}{R_0}\right) \right]^2
\over
\left[\left(1-\frac{2m}{R_0}\right)\left(1-\frac{a^2}{R_0^2}\right)\right]^{3/2}}
+
{ \frac{2m}{R_0}\left(1-\frac{a^2}{R_0^2}\right) + \frac{3a^2}{R_0^2}\left(1-\frac{2m}{R_0}\right) - \frac{4m a^2}{R_0^3} 
\over
\sqrt{\left(1-\frac{2m}{R_0}\right)\left(1-\frac{a^2}{R_0^2}\right)}} 
   \right\} ,
   \label{stablesymmass}
\end{eqnarray}
 and
 \begin{eqnarray}
R_0^2 \left[4\pi R_0^2\, \Xi_0 \right]''
&\geq & 
\frac{2a^2 }{\left( R_0^2-a^2 \right)} 
\left\{
-\frac{3 R_0^2(2R_0^2+a^2)}{\left( R_0^2-a^2 \right)^2} \sqrt{\left(1-\frac{2m}{R_0}\right)\left( 1-\frac{a^2}{R_0^2} \right)} 
+
\frac{4 R_0^2}{\left( R_0^2-a^2 \right)} 
{ \frac{m}{R_0}\left(1-\frac{a^2}{R_0^2}\right) + \frac{a^2}{R_0^2}\left(1-\frac{2m}{R_0}\right)
\over
\sqrt{\left(1-\frac{2m}{R_0}\right)\left(1-\frac{a^2}{R_0^2}\right)}}  
	\right.
	\nonumber \\
&& \qquad \qquad \qquad \left. 
  +
{\left[ \frac{m}{R_0}\left(1-\frac{a^2}{R_0^2}\right) + \frac{a^2}{R_0^2}\left(1-\frac{2m}{R_0}\right)\right]^2
\over
\left[\left(1-\frac{2m}{R_0}\right)\left(1-\frac{a^2}{R_0^2}\right)\right]^{3/2}}  
+
{ \frac{2m}{R_0}\left(1-\frac{a^2}{R_0^2}\right) + \frac{3a^2}{R_0^2}\left(1-\frac{2m}{R_0}\right) - \frac{4m a^2}{R_0^3}
\over
\sqrt{\left(1-\frac{2m}{R_0}\right)\left(1-\frac{a^2}{R_0^2}\right)}} 
 \right\}
\,,
\label{stablesymXi}
\end{eqnarray}
respectively.

\begin{figure}[!h]
\includegraphics[scale=0.50]{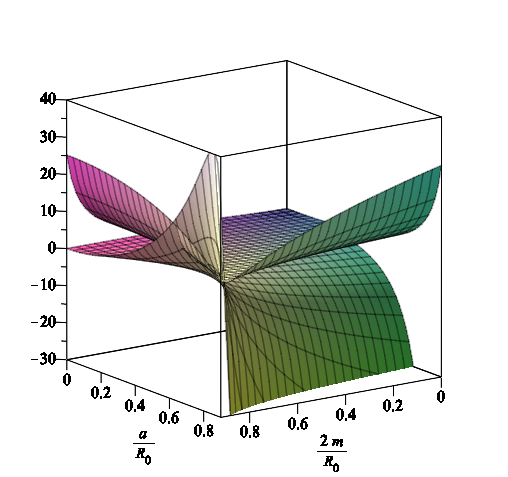}
\caption{The upper surface depicts the quantity $F_{4}(R_0,m,a)= R_0^2 \left[m_s(R_0)/R_0\right]''  $, and the stable region lies above that surface. 
On the other hand, the function $G_{4}(R_0,m,a)=R_0^2 \left[4\pi R_0^2\, \Xi_0 \right]''$ is depicted by the lower surface, and the stable region also lies above that surface. 
The final stability region of the solution lies above the intersection of both surfaces. 
{Note that inequality (\ref{stablesymXi}) serves to decrease the stability region for high values of $x$ and $y$}. 
See the text for more details.}
\label{fig:stable2}
\end{figure}

It is useful to express inequality (\ref{stablesymmass}) in the following dimensionless form $F_{4}(R_0,m,a)= R_0^2 \left[m_s(R_0)/R_0\right]''  $, and inequality (\ref{stablesymXi}) as $G_{4}(R_0,m,a)= R_0^2 \left[4\pi R_0^2\, \Xi_0 \right]''$. Both surfaces are depicted in figure \ref{fig:stable2}, and the final stability region is situated above the intersection of the surfaces. 

Note that we have considered the definition $x=2m/R_0$ for convenience, so as to bring infinite $R_0$ within a finite region of the plot. That is,  $R_0 \rightarrow \infty$ is represented as $x\rightarrow 0$; and $R_0 = 2m$ is equivalent to $x=1$. Thus, the parameter $x$ is restricted to the range $0<x<1$. We also define the
parameter $y=a/R_0$, which also lies in the range $0<y<1$.
It is interesting to note that the inequality (\ref{stablesymXi}) serves to decrease the stability region for high values of $x$ and $y$, as is transparent from figure \ref{fig:stable2}.
%=============================================================
%=============================================================

%=============================================================
\subsection{Two specific asymmetric cases}\label{SS3b:asymmetric}
%=============================================================

%=============================================================
\subsubsection{$a_{+} \neq a_-$ and $m_{+}=m_-$}\label{SS3b:asymmetric1}
%=============================================================

\begin{figure}[!h]
\includegraphics[scale=0.50]{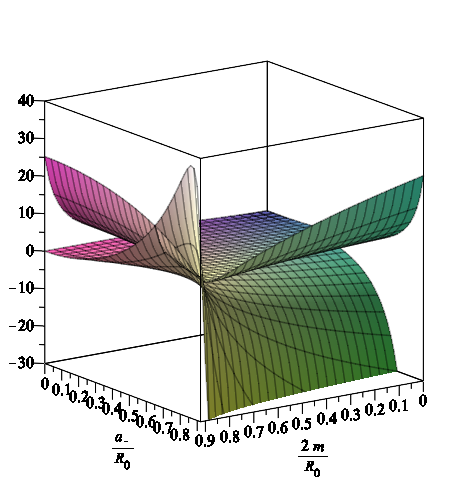}
\includegraphics[scale=0.50]{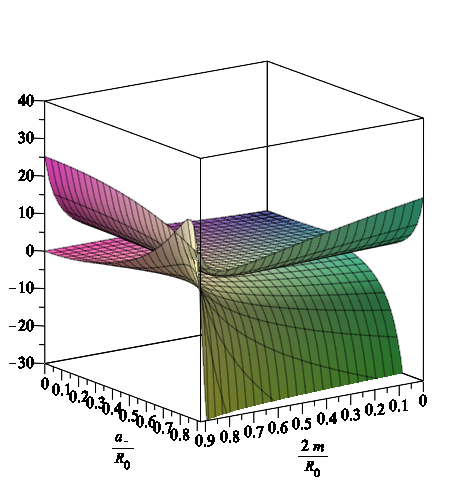}
\caption{Specific case of $a_{+} = \alpha\, a_-$ and $m_{+}=m_-$ for $\alpha<1$: The upper surface depicts the quantity $F_{5}(R_0,m,a_{\pm})= R_0^2 \left[m_s(R_0)/R_0\right]''  $, and the function $G_{5}(R_0,m,a_{\pm})=R_0^2 \left[4\pi R_0^2\, \Xi_0 \right]''$ is depicted by the lower surface. The stable regions are given above the surfaces. We have considered $\alpha=0.9$ in the left plot and $\alpha=0.4$ in the right plot. Note that decreasing the value of $\alpha$, serves to decrease the lower surface representing inequality (\ref{stability_constraint2}), and thus increase the final stability region. See the text for more details.}
\label{fig:stable5a}
\end{figure}

\begin{figure}[!h]
\includegraphics[scale=0.430]{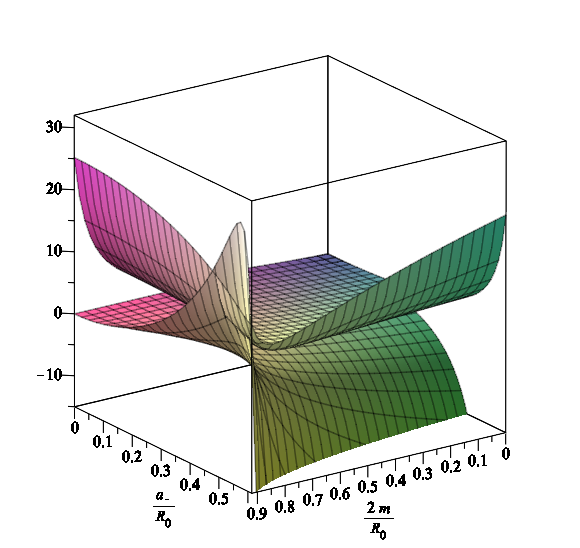}
\includegraphics[scale=0.430]{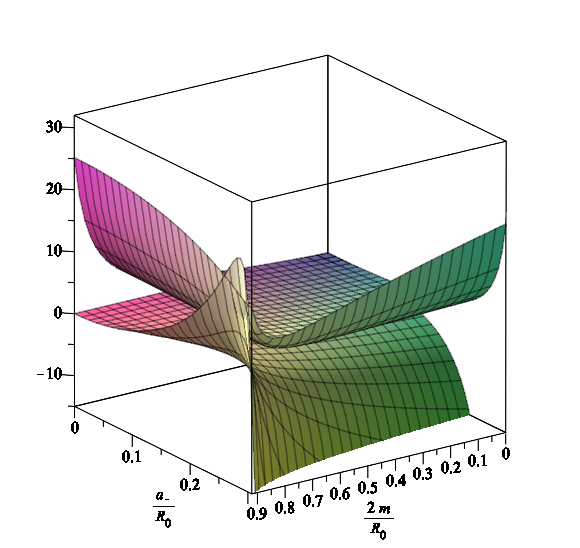}
\caption{Specific case of $a_{+} = \alpha\, a_-$ and $m_{+}=m_-$ for $\alpha>1$: The upper surface and lower surfaces depict the quantities $F_{5}(R_0,m,a_{\pm})= R_0^2 \left[m_s(R_0)/R_0\right]''  $ and $G_{5}(R_0,m,a_{\pm})=R_0^2 \left[4\pi R_0^2\, \Xi_0 \right]''$, respectively. The stable regions are given above the surfaces. We have considered $\alpha=1.5$ in the left plot and $\alpha=3$ in the right plot.
Note that increasing the value of $\alpha$, serves to decrease the lower surface representing inequality (\ref{stablesymXi}), and thus increase the final stability region. However, the range for $y$ decreases for increasing values of $\alpha$ (for $\alpha=1.5$, the range is $0<y<2/3$ and for $\alpha=3$, it is $0<y<1/3$). See the text for more details.}
\label{fig:stable5b}
\end{figure}

Consider the case of symmetric masses $m_{\pm}=m$, but with asymmetric bounce parameters $a_{+} \neq a_-$. In order to analyse the stability regions we define $a_{+} = \alpha\, a_-$, where $\alpha \in \Re^{+}$. Rather than write down the explicit form of the inequalities which are rather lengthy and messy, we present the dimensionless form of inequalities (\ref{stability_constraint1}) and (\ref{stability_constraint2}) as the upper and lower surfaces in the plots of figures \ref{fig:stable5a} and \ref{fig:stable5b}. Note that the specific case of $\alpha=1$ reduces to the analysis of the mirror symmetry considered in the previous subsection \ref{SS3a:symmetric}. We consider the dimensionless parameters $x=2m/R_0$, and $y=a_{-}/R_0$, in order to analyse the stability regions. In the following we separate the cases  $\alpha<1$ and $\alpha>1$.

\begin{itemize}
%%%
\item  
{\bf Specific case of $\alpha<1$}: 
the upper surface and lower surfaces in figure~\ref{fig:stable5a} depict the functions $F_{5}(R_0,m,a_{\pm})= R_0^2 \left[m_s(R_0)/R_0\right]''  $ and $G_{5}(R_0,m,a_{\pm})=R_0^2 \left[4\pi R_0^2\, \Xi_0 \right]''$, respectively, and the stability regions representing the inequalities (\ref{stability_constraint1}) and (\ref{stability_constraint2}) are  the regions above these surfaces.
The range of the dimensionless parameters is $0< x,y <1$. We have considered $\alpha=0.9$ in the left plot and $\alpha=0.4$ in the right plot of figure~\ref{fig:stable5a}. 
As the final stability region of the solution lies above the intersection of both surfaces, we note that decreasing the value of $\alpha$ qualitatively serves to decrease the lower surface representing inequality (\ref{stability_constraint2}), and thus increase the final stability region. 
This is transparent for high values of $x$ and $y$.
%%%
\item 
{\bf Specific case of $\alpha>1$}: 
the analysis is analogous to the above case and is depicted in figure~\ref{fig:stable5b}, however, here the left plot is given by $\alpha=1.5$ and the right plot by $\alpha=3$. The range of the dimensionless parameters is given by $0< x <1$ and $0< y <1/\alpha$.
As in the previous example, the final stability region of the solution lies above the intersection of both surfaces depicted in figure~\ref{fig:stable5b}. Note that increasing the value of $\alpha$, serves to decrease the lower surface representing inequality (\ref{stability_constraint2}), and thus increase the final stability region. However, the range for $y$ decreases for increasing values of $\alpha$ (for $\alpha=1.5$, the range is $0<y<2/3$ and for $\alpha=3$, it is $0<y<1/3$). This analysis is transparent for high values of $x$ and $y$ in figure \ref{fig:stable5b}.
\end{itemize}

%=============================================================
%=============================================================
%
%
%=============================================================
%=============================================================

%=============================================================
\subsubsection{$a_{+} = a_-$ and $m_{+} \neq m_-$}\label{SS3b:asymmetric2}
%=============================================================

\begin{figure}[!h]
\includegraphics[scale=0.430]{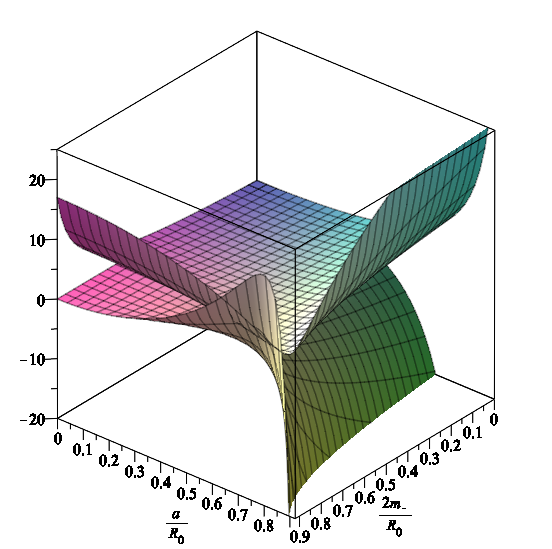}
\includegraphics[scale=0.430]{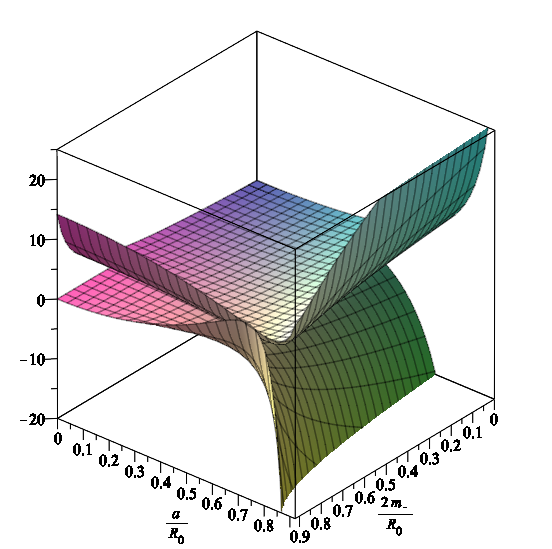}
\caption{Specific case of $a_{+} = a_-$ and $m_{+} = \alpha m_-$ for $\alpha<1$. We consider the dimensionless parameters $x=2m_{-}/R_0$, and $y=a/R_0$. We have considered $\alpha=0.9$ in the left plot and $\alpha=0.7$ in the right plot. Note that decreasing the value of $\alpha$, serves to decrease the lower surface representing inequality (\ref{stability_constraint2}), and thus increase the final stability region. See the text for more details.}
\label{fig:stable7a}
\end{figure}

\begin{figure}[!h]
\includegraphics[scale=0.430]{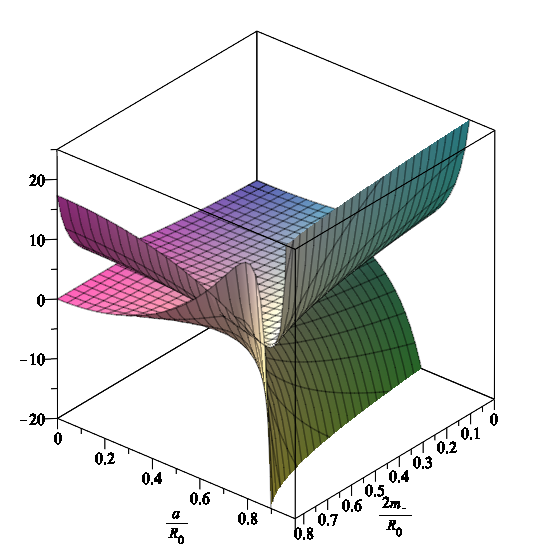}
\includegraphics[scale=0.430]{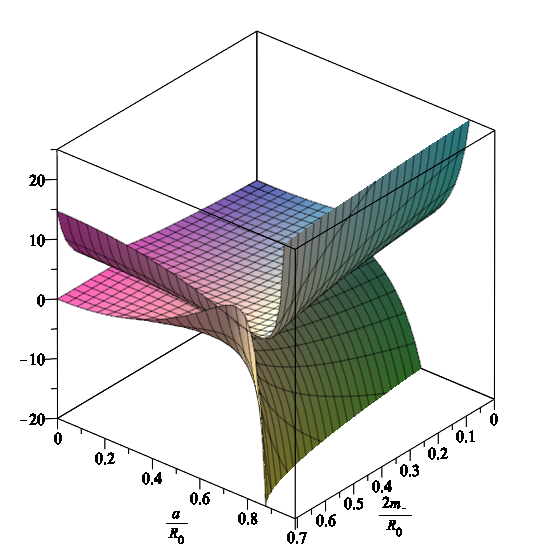}
\caption{Specific case of $a_{+} = a_-$ and $m_{+} = \alpha m_-$ for $\alpha>1$. We consider the dimensionless parameters $x=2m{-}/R_0$, and $y=a/R_0$. We have considered $\alpha=1.1$ in the left plot and $\alpha=1.3$ in the right plot. Note that increasing the value of $\alpha$, serves to decrease the lower surface representing inequality (\ref{stability_constraint2}), and thus increase the final stability region. See the text for more details.}
\label{fig:stable7b}
\end{figure}

Here we consider the case of asymmetric masses $m_{\pm} \neq m$, but with symmetric bounce parameters $a_{\pm} = a$, and for simplicity define $m_{+} = \alpha\, m_-$, where $\alpha \in \Re^{+}$. As in the specific case previously given above, we present the dimensionless form of inequalities (\ref{stability_constraint1}) and (\ref{stability_constraint2}) as the upper and lower surfaces in the plots of figures \ref{fig:stable7a} and \ref{fig:stable7b}. We consider the dimensionless parameters $x=2m_{-}/R_0$, and $y=a/R_0$, in order to analyse the stability regions, and as in the previous example we separate the cases  $\alpha<1$ and $\alpha>1$.

\begin{itemize}
%%%
\item  {\bf Specific case of $\alpha<1$}: the functions $F_{6}(R_0,m,a_{\pm})= R_0^2 \left[m_s(R_0)/R_0\right]''  $ and $G_{6}(R_0,m,a_{\pm})=R_0^2 \left[4\pi R_0^2\, \Xi_0 \right]''$ are depicted by the upper surface and lower surfaces in figure~\ref{fig:stable7a}, respectively, and the stability regions representing the inequalities (\ref{stability_constraint1}) and (\ref{stability_constraint2}) are given above these surfaces.
The range of the dimensionless parameters is $0< x,y <1$. We have considered $\alpha=0.9$ in the left plot and $\alpha=0.7$ in the right plot of figure~\ref{fig:stable7a}. 
The final stability region of the solution lies above the intersection of both surfaces. Note that decreasing the value of $\alpha$, serves to decrease the lower surface representing inequality (\ref{stability_constraint2}), and thus increase the final stability region.
%%%
\item {\bf Specific case of $\alpha>1$}: analogously to the above case, the stability regions are depicted in figure~\ref{fig:stable7b}. The left plot is given by $\alpha=1.1$ and the right plot by $\alpha=1.3$. The range of the dimensionless parameters is given by $0< x <1$ and $0< y <1/\alpha$.
As in the previous example, the final stability region of the solution lies above the intersection of both surfaces depicted in \ref{fig:stable7b}. Note that increasing the value of $\alpha$, serves to decrease the lower surface representing inequality (\ref{stability_constraint2}), and thus increase the final stability region. However, the range for $y$ decreases for increasing values of $\alpha$.

\end{itemize}

%=============================================================
%=============================================================

%=============================================================
%=============================================================

%=============================================================
\section{Conclusion}\label{S:conclusion}
%=============================================================

In this article we have considered thin-shell wormholes based on the recently introduced  black-bounce spacetimes. Specifically, by matching two black-bounce spherically symmetric spacetimes using the cut-and-paste procedure, we have analyzed the stability and evolution of dynamic thin-shell black-bounce wormholes. We have explored the parameter space of various models depending on the bulk masses $m_\pm$, and the values of the bulk bounce parameter $a_\pm$, investigating the internal dynamics of the thin-shell connecting the two bulk spacetimes, and demonstrating the existence of suitable stability regions in parameter space.
Several of these models are particularly useful in order to emphasize the specific features of these black bounce spacetimes. For instance, some examples are important to assess the simplifications due to symmetry between the two asymptotic regions, while other cases are useful to understand the asymmetry between the two universes used in traversable wormhole construction. 
Indeed, the interesting physics is encoded in the parameter $a_{\pm}$ which characterizes the scale of  the bounce, and that emphasizes the features of these spacetimes. In fact, the presence of the parameter $a_{\pm}$ induces the flux term in the conservation identity, which is responsible for the net discontinuity in the bulk momentum flux which impinges on the shell. Thus, due to this flux term, the bounce parameter places an additional constraint on the stability analysis of the spacetime geometry. 

From the linearized stability analysis one may assess and understand the (a)symmetry between the two universes in the traversable wormhole construction. For instance, consider the simple case of vanishing mass $m_{\pm}=0$ and $a_{\pm} \neq 0$ analysed in subsection \ref{SS2a:symmetric}, where large stability regions exist for $a_{\pm} \ll R_0$. However, the low stability region for the specific case of $a_+ \sim R_0$, increases significantly for the asymmetric case of decreasing the value of $a_-$. Another interesting example is the asymmetric case analysed in subsection \ref{SS2:asymmetric}, where we considered $a_+=0$ and $m_-=0$. Here, the low stability regions for $a_{-} \sim R_0$ and $m_{+} \sim R_0$ may be significantly increased by decreasing the asymmetric parameters, until large stability regions are present for $a_- \ll R_0$ and $m_+ \ll R_0$. 
The mirror symmetric case of $a_{\pm}=a$ and $m_{\pm}=m$, considered in subsection \ref{SS3a:symmetric} is of particular interest. Despite the fact that large stability regions exist for $a \ll R_0$ and $m \ll R_0$, the region at $a \approx R_0$ and $m \approx R_0$ is of particular interest. Here the flux term constraint kicks in and lowers the stability region significantly. In fact, analysing this specific region for the asymmetric case is particularly interesting, as one may explore the asymmetry between the universes in the wormhole construction by considering the case of $a_+ \neq a_-$ in subsection \ref{SS3b:asymmetric}. More specifically, by varying the relative values of $a_+$ and $a_-$, the analysis showed that one could increase or decrease the stability regions, and one may assess and understand the (a)symmetry between the two universes in the traversable wormhole construction.
In concluding, the constructions considered in this work are sufficiently novel to be interesting, and sufficiently straightforward to be tractable.

%\clearpage
\bigskip
%=====================================================
%=====================================================
\acknowledgments
%=====================================================
\noindent
FSNL acknowledges funding from the research grants No. UID/FIS/04434/2019, No. PTDC/FIS-OUT/29048/2017, and CEECIND/04057/2017.\\
AS acknowledges financial support via a PhD Doctoral Scholarship provided by Victoria University of Wellington.
AS is also indirectly supported by the Marsden fund, administered by the Royal Society of New Zealand.\\
MV was directly supported by the Marsden Fund, via a grant administered by the Royal Society of New Zealand.

%=====================================================

%=============================================================
\end{document}